\documentclass[prl,showpacs,draft]{revtex4}
\begin{document}
\title{Magnetism and half-metallicity at the O surfaces of ceramic oxides}
\author{S. Gallego}
\author{J.I. Beltr\'an}
\author{J. Cerd\'a}
\author{M.C. Mu\~noz}
\affiliation{Instituto de Ciencia de Materiales de Madrid, Consejo Superior de
Investigaciones Cient{\'{\i}}ficas,\\ Cantoblanco, 28049 Madrid, Spain}
 
\date{initially submitted 9 February}

\begin{abstract}
The occurence of spin-polarization at ZrO$_{2}$, Al$_{2}$O$_{3}$ and MgO
surfaces is
proved by means of \textit{ab-initio} calculations within the
density functional theory. Large spin moments, as high as 1.56 $\mu_B$, 
develop at O-ended  polar terminations, transforming the non-magnetic insulator into 
a half-metal. The magnetic moments mainly reside in the surface oxygen atoms 
and their origin is related to the existence of $2p$ holes of well-defined spin
polarization at the valence band of the ionic oxide. 
The direct relation between magnetization and local loss of donor charge makes
possible to extend the magnetization mechanism beyond surface properties.

\noindent
\end{abstract}

\maketitle

%%%%%%%%%%%%%%      

When dimensions are reduced to the nanoscale, we are faced to a new
understanding of the physical properties of matter: bulk insulators and
semiconductors exhibit metallic surfaces \cite{metsurf}, non-magnetic
materials get spin polarization when forming nanoparticles \cite{au}, 
unstable bulk structures exist in ultrathin film form \cite{metastable,noguera}, etc.
At the origin of these phenomena are the reduced dimensionality and the
enhanced role of the surfaces or boundaries in the final properties of the
system. Together with its inherent fundamental interest, this has important
technological consequences, auspicating the birth of new technologies
\cite{spintronics,magnetoelectronics}.
Special attention is devoted to magnetic low dimensional structures,
in particular as sources of spin current in the emerging field of
spintronics \cite{spintronics}.

In this letter, we report on the existence of large magnetic moments and 
half-metallicity at the O-rich surfaces of ceramic oxides, focusing on
ZrO$_{2}$, Al$_{2}$O$_{3}$ and MgO.
These are non-magnetic ionic insulators widely applied in bulk and thick film form, 
that have also been grown as ultrathin films and nanometric grains \cite{nanozro2}.
Their electronic structure can be roughly described as a valence band formed by 
the filled O $2p$ orbitals and a conduction band formed by the
empty metal levels. 
When a M$_{x}$O$_{y}$ unit -M being the metal donor and \textit{x,y} 
accounting for the particular metal to oxygen ratio- 
is broken to form the surface, the
loss of coordination of the surface O atoms originates
$2p$ holes in the valence band of the oxide. Our results show that
this generates high spin moments at the topmost O layer,
which induce magnetization at the adjacent planes and, remarkably, alter the
electronic structure of the oxide from insulating to half-metallic.

Very recently unexpected ferromagnetism has been measured in thin films
of undoped non-magnetic oxides, like HfO$_2$ and ZrO$_2$,
possibly assigned to the presence of lattice defects concentrated at the film interface. \cite{hfo2,aps}.
Here we prove the existence of a magnetization mechanism rooted in 
the loss of donor charge of the O atoms, something that can also occur in films
with cation vacancies. 
This provides an explanation for the origin of the magnetic moments
of these so called 'd-zero' ferromagnets. Furthermore, we predict that this kind
of magnetism can be extended to a wider
class of non-magnetic oxides, like MgO and Al$_2$O$_3$, where the cation is not 
necessarily a $d$ metal.

The presence of defects has already been found to transform certain bulk ionic
materials from insulating non-magnetic to ferromagnetic:
either by adding transition metal dopants \cite{tio2}, as occurs with TiO$_{2}$,
SnO$_{2}$ and ZnO, or by a kind of molecular Hund's rule coupling for the
solid state \cite{hund-sol}, as in the case of dilute divalent Ca 
vacancies in CaO and dilute neutral B$_6$ vacancies in the hexaborides.
\cite{hexaborides}.
Although the magnetization mechanism
reported here has an electrostatic origin similar to the one proposed in bulk CaO,
new ingredients are present: the localization of the spin moments close to
the surface, and the extended character of the polarized $2p$ valence band states.
This suggests that intrinsically low-dimensional structures of the oxides considered,
such as grains or ultrathin films, can be magnetic.

Our results are obtained from first-principles spin-polarized calculations
within the DFT (density functional theory) under the GGA (generalized gradient 
approximation) for exchange and correlation \cite{gga}.
For highly correlated systems such as transition metal oxides (TMO), this
approach predicts too small magnetic moments and inaccurate magnetic energies
due to its failure in describing the localized nature of the $d$ electrons \cite{dftu}.
However, this is not the situation for the ceramic oxides considered here,
with a valence band formed by O $2p$ orbitals. In fact,
there are no valence $d$ electrons in MgO and Al$_2$O$_3$, while it is widely 
recognized the success of the DFT-GGA formalism, with and without self-energy corrections 
\cite{louie}, to determine the structural and electronic properties of ZrO$_2$,
which are in excellent agreement with experiments \cite{spectra}.
We use the SIESTA package \cite{siesta} with basis sets formed by double-zeta polarized 
localized numerical atomic orbitals. More details about the conditions of the 
calculations can be found elsewhere \cite{prb-zro2}. 

We have studied binary insulating oxides of 
different crystal structures: MgO, Al$_2$O$_3$ and ZrO$_2$,
for which few calculations have considered the possibility
of spin polarization at clean surfaces or in defective structures\cite{noguera}.
MgO adopts the rock-salt lattice, the corresponding (111) surfaces consisting of 
a stacking of alternated pure Mg and O planes.
$\alpha$-Al$_2$O$_3$ can be viewed as an hexagonal close-packed array of
O atoms with the Al atoms occupying two thirds of the interstitial octahedral 
sites. Along the [0001] direction, consecutive O planes are separated by two Al layers.
ZrO$_2$ displays three polymorphs at room temperature \cite{mzro2}, either in
its pure form (m-ZrO$_2$, monoclinic), or by dopage with substitutional
cations (t-ZrO$_2$, tetragonal, and c-ZrO$_2$, cubic). The cubic
structure corresponds to the CaF$_2$ lattice, which along the (111) direction
leads to three possible terminations, labelled according to the composition
of the two topmost layers: {\it O-O-}, {\it O-Zr-} and {\it Zr-O-}. t-ZrO$_2$ can be
obtained from the cubic unit cell elongating one of the equivalent edges of
the cube and introducing small opposite shifts at the O sublattice
positions. The m-ZrO$_2$ lattice is more complex, but along the [001]
direction it defines a layered structure of chemical stacking sequence
O-O-Zr. Here we have considered all low-index terminations of c-ZrO$_2$, and
selected t-ZrO$_2$, m-ZrO$_2$, MgO and Al$_2$O$_3$ surfaces. 
Each surface is modelled as a
periodic slab containing a vacuum region of 10 $\AA$. The width of the
vacuum region is enough to inhibit interaction between neighbouring
surfaces. The number of M$_{x}$O$_{y}$ units in a slab depends on the particular
structure, and ranges from 5 to 7. The slabs are symmetric about the central
plane, to avoid an artificial divergent Madelung energy, 
although we have verified that all the conclusions of this work remain unaltered
if asymmetric slabs are considered instead.
Bulk-like behaviour is attained at the innermost central layers. The atomic positions
are allowed to relax until the forces on the atoms are less than 0.06 eV/$\AA$. 
For comparison, a non-spin polarized (NSP) calculation is also performed for
each case. 
The energy reduction due to spin polarization for all surfaces with a
magnetic ground-state is shown in Table \ref{tener}. 
It has been obtained as the total energy difference between the relaxed SP and
NSP slabs. The large values in the table are in the range of those obtained for magnetic 
bulk oxides \cite{dftu}, and should not be taken as an exact prediction of the
actual values, but as a correct description of the order of magnitude and global trend.

Table \ref{tmagmo} summarizes our first main result. 
It shows the total magnetic moments at the surface O plane
for all the structures which present a magnetic ground state, together with 
their in-plane ($||$) and normal ($\perp$) $p$ projections onto the surface.
All spin moments arise almost exclusively from the $p$ electrons
at the valence band of the oxide.
They reach large values, ranging from 0.8 to 1.6 $\mu_B$.
Notice that only the O-rich oxide surfaces are magnetic. 
Both the cation-ended and the non-polar O surfaces
not in the table do not show any spin polarization.
The distinct characteristic of all these magnetic structures is that
they are polar divergent surfaces \cite{tasker} with a loss
of cationic coordination for the outermost O atoms. This already suggests that
the O spin polarization is intimately related to the decrease of the 
oxygen ionic charge at the surface with respect to the bulk. 
The correlation is evident after inspection of Table \ref{tmagmo}, where the 
rightmost columns provide the O Mulliken populations at the surface
layer and the bulk. In all these cases the surface O lose
ionic charge approaching a charge neutral state. 
Moreover, the smaller the oxygen ionic charge, the larger its associated magnetic moment. 

Indeed, for the calculated surfaces with a non-magnetic ground state we obtain O 
charges close to the bulk ones. 
An example is given in table \ref{t111}, which compares the layer resolved Mulliken populations 
and moments for the two O-ended c-ZrO$_2$(111) surfaces.
>From the results for c-ZrO$_2$(111)$_{O-O-}$ in table \ref{t111},
we can again observe the clear correlation between the
reduction of the ionic charge and the size of the magnetic moment.
This means that the spin polarization is a response of the system to the loss
of transferred charge.
Even if the magnetization is a local effect, rooted in
the lack of donor electrons for the outermost O atoms, the interaction of these
atoms with the lower adjacent planes cannot be neglected:
the spin polarization at the surface layer
induces a magnetic moment at the neighbouring planes, which decays as we deepen into the bulk.

Our second main result is the finding that all magnetic surfaces are half-metallic. 
Half-metals are strong candidates for the development of spintronic devices, as they can naturally 
produce electronic currents of well-defined spin polarization \cite{halfmetals}.
Figure \ref{fdos} shows the density of states (DOS) at the three uppermost layers and
the bulk for two representative Al$_2$O$_3$(0001) and c-ZrO$_2$(001) surfaces,
both for SP (left) and NSP (right) calculations. 
In all the magnetic structures investigated, the Fermi
level (E$_F$) crosses the valence band of the topmost layer, although
some surfaces present a very localized minimum at E$_F$, as occurs in Al$_2$O$_3$(0001).
Remarkably, the bands crossing E$_F$ have always a well
defined spin polarization, the majority spin states being full while the
minority spin band is partially filled. The result, common to all the magnetic
surfaces studied here, is a half-metallic system.

The differences between the magnetic surfaces manifest in the details of the
electronic distributions. This can be best observed regarding the shape of the DOS
in figure \ref{fdos}.
For c-ZrO$_2$(001), the DOS at the topmost O plane is bulk-like except
for the usual surface narrowing, both for the SP and NSP calculations.
The effect of the spin exchange is mostly the shift of the minority
spin band. At the SP Al$_2$O$_3$(0001) surface, the most significant effect is the 
reduction of the DOS at E$_F$, something 
also observed at c-ZrO$_2$(111)$_{O-O-}$ and MgO(111). For these surfaces, the 
shape of the majority spin DOS is almost bulk-like, while a deep redistribution of 
charge affects the minority spin electrons.

The major changes observed in the minority spin DOS are also present in the spatial
charge distributions. This can be directly seen in the left and central panels of figure \ref{fchdens}, 
which depict the spin resolved charge density differences (CDD, total charge density
minus the superposition of atomic charge densities) of the MgO(111) and c-ZrO$_2$(111)$_{O-O-}$
surfaces, within a plane normal to the surface along the bond direction between 
a surface O atom and a nearest neighbour at the layer below.
The majority spin band being completely filled, as was shown in the DOS,
the corresponding CDD adopts the spherical symmetry of the bulk. 
On the contrary, anisotropic lobular shapes appear in the minority charge. 
This assymetry between the spatial distributions of the 
spin states is a common feature to all the magnetic structures studied here, although 
the effect is specially relevant for the surfaces shown in figure \ref{fchdens}. 
The rightmost panels of the figure show the corresponding spin density difference 
(SDD), CDD of majority spin minus CDD of minority spin. 
The crystal field induces a high directionality of the spin moments, involving orbitals
along specific directions. For the c-ZrO$_2$(111)$_{O-O-}$ and MgO(111) systems
this leads to a highly anisotropic distribution of the magnetic charge, also evidenced 
in the m$_{||}$ and m$_{\perp}$ projections of table \ref{tmagmo}. 
The specific dominant distribution depends on the particular
structure: contained within the surface plane for c-ZrO$_2$(111)$_{O-O-}$ and
normal to the surface for MgO(111).

Finally, some considerations about energetics should be mentioned.
To analyze the surface energetics requires to consider all mechanisms which stabilize
the surfaces under study \cite{noguera,dulub}, a huge task beyond
the scope of this work.
Nevertheless, it is well known that bulk metastable terminations coexist in grain 
boundaries, and they even become energetically favourable in 
ultrathin films and nanostructures \cite{nanoE}.
Based on thermodynamic arguments \cite{stability, eichler}, the surface energy of the polar
surfaces under study may be lowered below that of the corresponding non-polar terminations
under high O pressures when the spin polarization is included.
On the other hand, we have also performed preliminar studies of the effect 
of Zr vacancies in bulk ZrO$_2$ that 
confirm the existence of the magnetization mechanism presented here.
Similar conclusions have been reported for bulk HfO$_2$ in the presence of
Hf vacancies \cite{sanvito}. 

In conclusion, we have proved the existence of local magnetism at the ideal
O-ended polar surfaces of ceramic oxides. Large spin moments are formed at the
outermost surface plane due to the lack of donor charge and the subsequent
creation of $2p$ holes in the valence band of the oxide. This leads to a
half-metallic ground state, with the majority spin band
completely filled and only minority spin states at the Fermi level.
The orientation of the magnetic moment with respect to the surface plane
can be tuned selecting adequate crystal orientations.
The fact that polar surfaces of nonmagnetic materials with large bulk bandgap 
exhibit a 100 \% spin polarization at the Fermi level opens a new route to the 
manipulation of spin currents. 
In fact, the mechanism presented here can explain the unexpected magnetism measured 
in thin films of HfO$_2$ and ZrO$_2$ \cite{hfo2,aps}, relating it to the existence 
of cation defects. Even more, the present ability to produce
ultrathin films and nanometric grains of these materials make feasible the fine
control of their structure and stoichiometry, implying the possibility to extend
this surface magnetism to any low-dimensional configuration.

%%%%%%%%%%%%%%
This work has been partially financed by the Spanish Ministerio de
Educaci\'on y Ciencia and the DGICyT under contracts MAT2003-04278, MAT2004-05348
and CAM2004-0440, and by the Ram\'on y Cajal program. J.I. B. acknowledges
financial support from the I3P program of the CSIC.

%%%%%%%%%%%%%%

\begin{table}[tener]
\caption{Energy reduction (eV/M$_x$O$_y$ unit) of the SP relaxed slab with respect to the corresponding NSP case
for all magnetic surfaces under study.}
\label{tener}\centering \vspace{0.5cm}
\begin{tabular}{p{20mm}p{13mm}p{10mm}p{10mm}|p{10mm}|p{10mm}}
 \multicolumn{4}{c|}{ZrO$_2$} & \multicolumn{1}{c|}{Al$_2$O$_3$} & \multicolumn{1}{c}{MgO} \\
  c-(111)$_{O-O-}$  &  m-(001) & t-(001) & c-(001) & \multicolumn{1}{c|}{(0001)} & \multicolumn{1}{c}{(111)} \\
\hline 
\multicolumn{1}{c}{0.91} & \multicolumn{1}{c}{0.44}  & \multicolumn{1}{c}{0.22} & \multicolumn{1}{c|}{0.22} & \multicolumn{1}{c|}{1.37} & \multicolumn{1}{c}{0.44} \\
\end{tabular}%
\end{table}          

\begin{table}[tmagmo]
\caption{Spin moments (in $\mu_B$) at the topmost layer of the surfaces of Table \protect\ref{tener},
together with the decomposition of the $p$ orbital contribution
parallel and normal to the surface plane. The rightmost columns provide the corresponding Mulliken 
charge populations (Q) compared to the inner bulk value (Q$_b$).
The O$_1$ and O$_2$ entries for the non-cubic ZrO$_2$ structures refer to the two inequivalent
in-plane positions.}
\label{tmagmo}\centering \vspace{0.5cm}
\begin{tabular}{p{26mm}|p{1mm}p{8mm}p{8mm}p{9mm}p{8mm}p{8mm}}
Surface                 &  & {\bf m$_{tot}$}  & m$^p$$_{||}$ & m$^p$$_{\perp}$ & {\bf Q} & Q$_b$\\
\hline
c-ZrO$_2$(111)$_{O-O-}$ &  & {\bf 1.56}      & 1.50     & 0.04        & {\bf 6.02} & 6.79 \\
m-ZrO$_2$(001)-O$_1$    &  & {\bf 1.43}      & 0.81     & 0.61        & {\bf 6.10} & 6.66 \\
\hspace{18.5mm}-O$_2$   &  & {\bf 1.43}      & 0.81     & 0.60        & {\bf 6.11} &  \\
t-ZrO$_2$(001)-O$_1$    &  & {\bf 1.19}      & 0.54     & 0.64        & {\bf 6.13} & 6.76 \\
\hspace{17mm}-O$_2$     &  & {\bf 0.59}      & 0.55     & 0.03        & {\bf 6.17} &  \\
c-ZrO$_2$(001)          &  & {\bf 0.83}      & 0.62     & 0.21        & {\bf 6.29} & 6.79 \\
Al$_2$O$_3$(0001)       &  & {\bf 0.97}      & 0.74     & 0.23        & {\bf 6.10} & 6.49 \\
MgO(111)                &  & {\bf 0.83}      & 0.06     & 0.76        & {\bf 6.37} & 6.72 \\
\end{tabular}%
\end{table}          

\begin{table}[t111]
\caption{Layer resolved spin moments (in $\mu_B$) and Mulliken populations
for the O-ended c-ZrO$_2$(111) surfaces. The layers, labelled according to their composition,
are ordered from the surface (top row) to the bulk.}
\label{t111}\centering \vspace{0.5cm}
\begin{tabular}{p{9mm}|p{8mm}p{2mm}p{8mm}|p{8mm}p{5mm}}
Layer & \multicolumn{3}{c|}{{\it O-O-}} & \multicolumn{2}{c}{{\it O-Zr-}} \\
      & \multicolumn{1}{r}{m$_{tot}$} & & \multicolumn{1}{c|}{Q} & \multicolumn{1}{c}{m$_{tot}$} &  \multicolumn{1}{c}{Q} \\
\hline
\multicolumn{1}{c|}{O} & \multicolumn{1}{r}{1.56} & & \multicolumn{1}{c|}{6.02} &  &  \\
\multicolumn{1}{c|}{O} & \multicolumn{1}{r}{0.47} &  &\multicolumn{1}{c|}{6.57} & \multicolumn{1}{c}{--} & 6.68 \\
\multicolumn{1}{c|}{Zr} & \multicolumn{1}{r}{-0.13}  && \multicolumn{1}{c|}{2.60} & \multicolumn{1}{c}{--} & 2.49 \\
\multicolumn{1}{c|}{O} & \multicolumn{1}{r}{0.07} &  &\multicolumn{1}{c|}{6.79} & \multicolumn{1}{c}{--} & 6.79 \\
\end{tabular}%
\end{table}

%******* FIGURES
%\newpage
 
\begin{figure}
\caption{ \label{fdos}
DOS for the three topmost planes and the bulk of Al$_2$O$_3$(0001) and c-ZrO$_2$(001)
after SP (left) and NSP (right) calculations.
Positive (negative) values correspond to majority (minority) spin states. 
Energies are referred to the Fermi level.
The total spin moments at each layer are indicated for the SP case.}
\end{figure}
 
\begin{figure}
\caption{ \label{fchdens}
Left and middle panels: spin resolved CDD for the (top) MgO(111) and (bottom) 
c-ZrO$_2$(111)$_{O-O-}$ surfaces. Right panels: corresponding SDD.}
\end{figure}
 
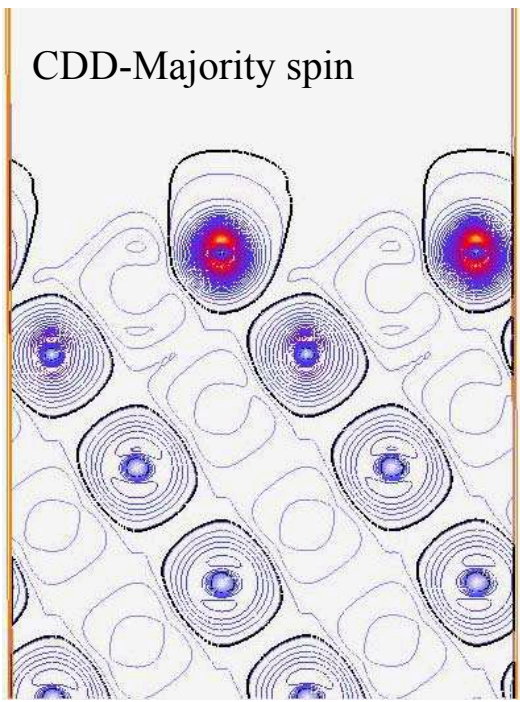
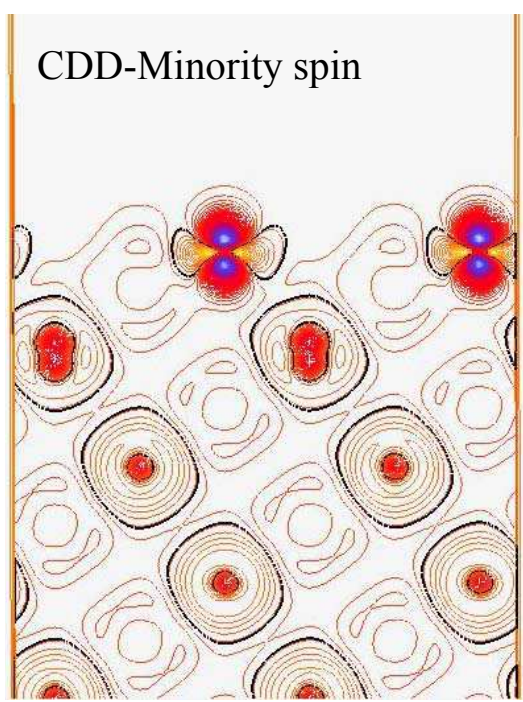
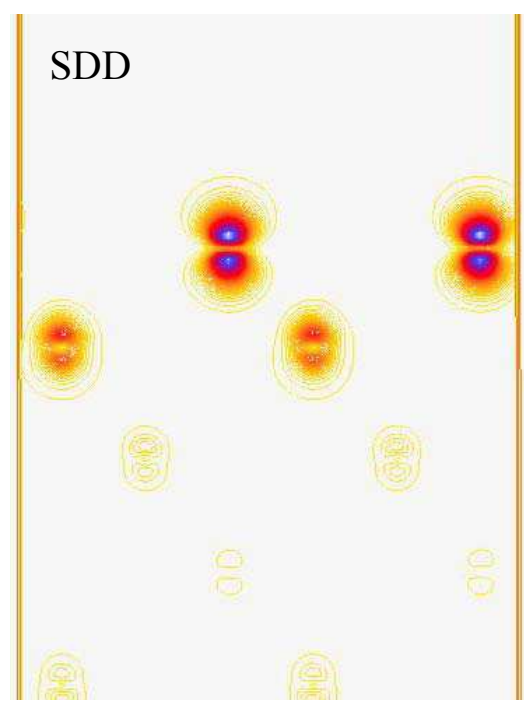
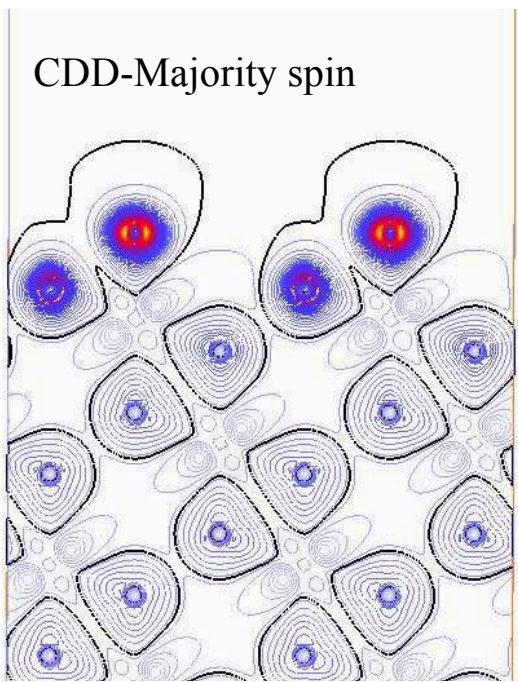
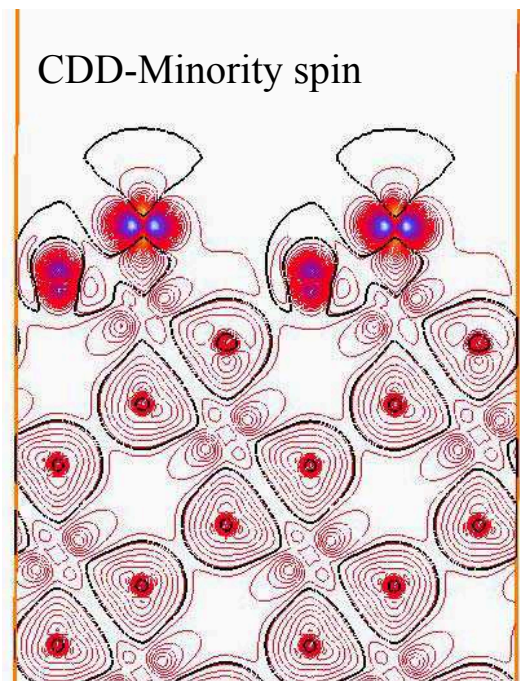
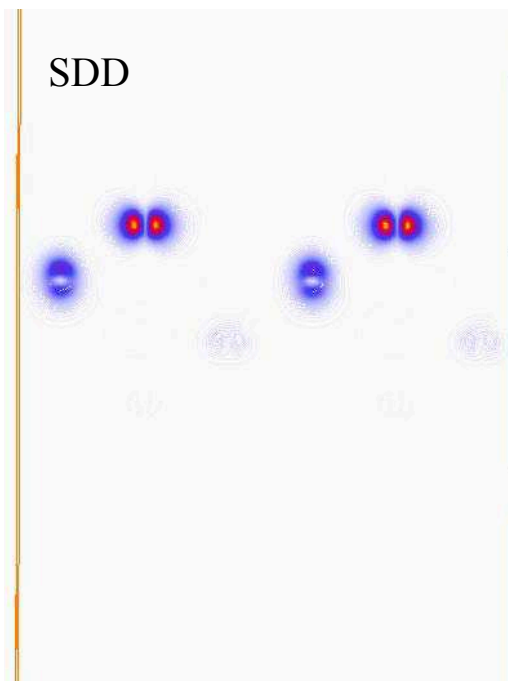
\end{document}